\documentclass[conference]{IEEEtran}
\IEEEoverridecommandlockouts
\usepackage{cite}
\usepackage{amsmath,amssymb,amsfonts}
\usepackage{algorithmic}
\usepackage{graphicx}
\usepackage{textcomp}
\usepackage{xcolor}
\usepackage{tikz}
\usetikzlibrary{shapes.geometric, arrows, positioning}
\usepackage{hyperref}
\usepackage{booktabs}
\usepackage{comment}
\def\BibTeX{{\rm B\kern-.05em{\sc i\kern-.025em b}\kern-.08em
    T\kern-.1667em\lower.7ex\hbox{E}\kern-.125emX}}
\begin{document}

\title{TheraQuest: A Gamified, LLM-Powered Simulation for Massage Therapy Training}

\author{\IEEEauthorblockN{1\textsuperscript{st} Shengqian Wang}
\IEEEauthorblockA{\textit{Faculty of Science} \\
\textit{Ontario Tech University}\\
Oshawa, Canada \\
Shengqian.wang@ontariotechu.net}
}

\maketitle

\begin{abstract}
Massage therapy training emphasizes hands-on techniques and effective therapist--patient communication. However, many educational programs struggle to provide realistic practice scenarios. To address this problem, we propose \textbf{TheraQuest}, a gamified, web-based simulation platform that employs large language models (LLMs) to generate diverse virtual patients with varying symptoms and cultural backgrounds. Through interactive dialogue, anatomical decision-making, and immediate assessment, trainees develop both diagnostic reasoning and empathetic communication skills in a low-risk environment. Unlike exclusively VR-based solutions, TheraQuest remains accessible via standard web browsers, mitigating the cost and discomfort associated with extended headset use. Preliminary testing suggests that integrating LLM-driven virtual patients with real-time skill metrics can enhance trainee engagement and help bridge the gap between theoretical knowledge and clinical proficiency. 
\end{abstract}


\section{Introduction}
Massage therapy has long been recognized as an effective method worldwide for addressing musculoskeletal problems, reducing stress, and promoting overall well-being \cite{ref1,ref2}. It serves a wide range of patients, particularly those in high-demand academic and work environments, such as IT professionals and heavy-duty laborers who develop posture-related issues \cite{ref92,ref93,ref94}, as well as individuals suffering from injuries (e.g., postoperative complications \cite{ref95,ref96,ref97} or sequelae \cite{ref98}) or conditions such as muscular torticollis \cite{ref90}, migraines \cite{ref99}, and cardiac defects \cite{ref100}. 

To become a qualified massage therapist, individuals must complete professional education and supervised practice to gain proficiency in anatomy, physiology, and hands-on techniques \cite{ref8, ref91}. In Canada, trainees must spend 18 to 36 months studying at accredited colleges and then complete 500 to 1000 hours of supervised practical training \cite{ref16,ref17} to become a Registered Massage Therapist (RMT). However, this traditional training approach may not fully address patient--therapist communication concerns, particularly in cases where patients come from diverse cultural or linguistic backgrounds. Therefore, massage therapists may need additional training hours if they struggle with diagnosing patient needs or adapting treatment approaches.

Effective communication between patients and therapists relies on precise verbal descriptions of symptoms. Canada, as a country with a diverse immigrant population, has patients with varying language and cultural backgrounds \cite{ref17}, which can sometimes hinder patient--therapist communication. For instance, patients without English or French proficiency may use simple words to describe complex conditions or passively agree with the therapist's assessments. As a result, therapists may misunderstand the patient's discomfort or its root causes and apply incorrect treatments.

On the other hand, the misconception that ``the harder, the better`` leads to various issues \cite{ref89}, particularly when some therapists apply excessive pressure in an attempt to provide immediate yet temporary relief. Unfortunately, this temporary relief worsens the patient's symptoms, damages the professional reputation of the therapist, and discourages patients from seeking further care. In such cases, patients may distrust the therapist or resort to painkillers rather than addressing the underlying issues.

To address these communication challenges, we designed \textbf{TheraQuest}, a gamified web-based training application that simulates real-world therapeutic massage scenarios. The system uses large language models (LLMs) to generate diverse virtual patients, each with unique symptoms and detailed profiles that closely mirror real-life conditions. These virtual patients incorporate both textual and verbal information, requiring trainees to determine optimal treatments, locate treatment areas, adjust pressure levels, and document key details within an interactive graphical user interface (GUI). Immediate feedback from patients, along with an expert assessment summary, enables trainees to refine their diagnostic skills, enhance communication strategies, and effectively translate theoretical knowledge into practical applications.

In this study, we begin by reviewing current massage therapist training approaches and exploring the use of gamified simulation applications in both general and medical education, including those specifically aimed at massage therapists and implementations of virtual patients. We then propose and evaluate our TheraQuest prototype, comparing it with other available applications to assess its effectiveness and potential impact.

\section{Related Work}
This section examines the current literature on traditional massage therapy training, the use of virtual patients for clinical education, and the growing trend of gamified simulation in health care, including its applications and gaps in the context of massage therapy.

\vskip 2mm
\noindent \textbf{Massage Therapy Training and Practice.}
Massage therapy education combines theoretical anatomy and physiology knowledge with the hands-on skills through clinical practice. 

Understanding how massage therapists are educated and practice is the first important step in identifying the gaps that new approaches must address. Moyer et al. \cite{ref3} reviewed 37 studies related to massage therapy efficacy, and the results indicated that massage therapy can address: 1). State anxiety, 2). Negative mood, 3). Immediate assessment of pain, 4). Cortisol levels, 5). Blood pressure, 6). Heart rate, 7). Trait anxiety, 8). Depression, and 9). Delayed assessment of pain. Although their findings support the therapeutic benefits of massage, the market still urgently need more competent practitioners who can tailor treatments to individual needs. Finally, they suggest developing new technologies or theoretical models to enhance the efficacy of massage therapy. They also emphasize the importance of improving therapist-patient communication in future research.

Sherman et al. \cite{ref4} took a closer look at the training and practice patterns of massage therapists in the United States, conducting a large-scale survey (N=226) that revealed significant variability in educational hours, practical experience, and ongoing professional development. Their findings suggest that, while core clinical competencies can be achieved, the depth of training in communication and diagnostic reasoning varies widely, potentially impacting how effectively therapists adapt treatments for culturally or linguistically diverse clients. This variability has significant implications for the United States practice environments, where standardized educational materials must accommodate diverse patient needs.

\vskip 2mm
\noindent \textbf{Virtual Patients in Healthcare Education.} Virtual patients, also known as computer-generated characters designed to mimic real patient behaviours, have gained attention as a way to standardize and improve clinical training across medical professions \cite{ref23}. Cook and Triola \cite{ref6} presented a critical literature review identifying virtual patients as valuable to refine diagnostic reasoning and communication skills in a risk-free environment. However, they also noted a lack of methodological consistency across different studies, calling for more robust research designs and clearer integration into health domains.

McCoy et al. \cite{ref5} conducted a comprehensive landscape review of gamification and multimedia in medical education, examining five electronic games, four mobile applications, and twelve virtual patient simulation tools. Their study identified seven key educational advantages of gamified training platforms: 1). Increased engagement, 2). Enhanced collaboration, 3). Real-world application, 4). Clinical decision-making, 5). Distance training, 6). Learning analytics, and 7). Swift assessment. They suggested that virtual patients, when combined with interactive elements such as immediate assessment or branching dialogue, can significantly enhance learner engagement and skill retention. However, the authors also noted that virtual patient solutions remain underutilized in health domains, including massage therapy.

\vskip 2mm
\noindent \textbf{Gamified Simulation for Education and Clinical Reasoning.} Beyond virtual patients alone, gamification, the integration of game design elements such as scoring, rewards, and challenges—has become an increasingly popular strategy to increase motivation and improve learning in educational fields \cite{ref26,ref27,ref28,ref30 }. Plackett et al. \cite{ref12} performed a systematic review of articles from 1990 to January 2022. They initially identified 8,186 articles, but only 19 met the inclusion criteria. The results demonstrate that gamified virtual patient tools can improve (11/19, 58\%) or significantly improve (4/19, 21\%) clinical reasoning skills among undergraduate medical students, although they stressed the importance of aligning tasks with specific learning objectives and providing ongoing assessment loops.

\vskip 2mm
\noindent \textbf{Simulations in Web-based Simulations.} 

Web-based simulations have been recognized as one of the most powerful and effective approaches for enhancing training in medical domains, especially in skill-based disciplines such as surgery and rehabilitation. Unlike traditional educational models that rely on textbooks or static video demonstrations, these new interactive platforms offer learners an immersive environment to practice hands-on skills. For example, Cowan et al. \cite{ref30} developed a web-based simulation for total knee arthroplasty training using real-time 3D visualization and interactive decision-making. Their study demonstrated that serious gaming environments could improve skill retention, engagement, and trainee confidence compared to conventional learning methods. The application of such simulations is particularly relevant in fields where hands-on practice is crucial but logistically constrained, such as massage therapy, where access to live patient scenarios is often limited. 

\vskip 2mm
\noindent \textbf{Gamified Simulation in Medical Education: Scope and Gaps.} Within the broader arena of health-professional training, augmented reality (AR), virtual reality (VR) and mixed reality (MR) platforms have captured particular attention. Barteit et al. \cite{ref11} systematically reviewed 27 articles with 956 study participants on head-mounted AR/VR devices in medical education, concluding that immersive applications can improve procedural training, situational awareness, and trainee satisfaction. Most of the studies (n=17, 63\%) found it effective, and some studies (n=4, 15\%) found it partially effective. However, the authors pointed out that randomized controlled trials remain relatively sparse and that cost or logistical barriers can impede widespread adoption.

\begin{table}[ht]
    \centering
    \caption{Popular AR/VR/MR Devices Ranked by Approximate Weight (Lightest to Heaviest)}
    \label{tab:arvrweights}
    \begin{tabular}{l l c}
    \toprule
    \textbf{Device} & \textbf{Type} & \textbf{Approx. Weight (g)} \\
    \midrule
    HP Reverb G2          & VR     & 500 \\
    Oculus Quest 2        & VR     & 503 \\
    Oculus Quest 3        & VR     & 510 \\
    PlayStation VR2       & VR     & 560 \\
    HTC Vive              & VR     & 563 \\
    Microsoft HoloLens 2  & AR/MR  & 566 \\
    Apple Vision Pro      & MR     & 650 \\
    Valve Index           & VR     & 809 \\
    \bottomrule
    \end{tabular}
\end{table}

\vskip 2mm
\noindent \textbf{Gamified Simulation in Massage Therapy and Related Disciplines.} Despite the growing body of work on virtual and gamified simulations in medicine and nursing, research focused on massage therapy remains very limited. Shema et al. \cite{ref7} provided one of the few studies (N=60) of a simulation-based approach specifically tailored to physical rehabilitation, incorporating a virtual reality treadmill program to improve gait, mobility, and postural control. Although their work targeted ambulatory physical therapy rather than massage therapy, the study demonstrated the feasibility of using virtual environments to simulate real-world patient challenges.

Expanding beyond physical therapy, Su and Cheng \cite{ref20} introduced a gamified rehabilitation system focused on creativity that uses an adaptive fuzzy inference model to dynamically adjust the difficulty of the task based on patient emotions. This approach, while not explicitly designed for massage therapy, illustrates how intelligently adaptive assessment can increase engagement and outcomes in therapeutic contexts. Similarly, Cortés-Pérez et al. \cite{ref21} tested Kahoot-based gamification for physiotherapy students, observing improved motivation and preliminary evidence of enhanced learning. Choo and May \cite{ref22} explored virtual mindfulness meditation using immersive technology, another innovative approach to health-oriented gamification that demonstrates how virtual platforms can extend beyond purely cognitive training into stress reduction and mind-body therapies. 

Although immersive applications have grown in popularity and technology has advanced rapidly in recent years, the physical burden of immersive devices remains a major challenge. The weights of popular AR/VR/MR devices are summarized in Table \ref{tab:arvrweights}. Requiring trainees to wear these devices for extended periods, like several hours can be impractical for training. As a result, web-based games continue to dominate gamified simulations and are likely to remain the preferred approach in the near future.

\vskip 2mm
\noindent \textbf{Toward a Gamified, Virtual-Patient Approach in Massage Therapy.} Taken together, these studies reveal a substantial gap in the use of gamified applications and virtual patient simulations that directly or indirectly relate to massage therapy. Although foundational research on massage therapy illustrates the importance of both hands-on technique and patient-centered communication \cite{ref3,ref4}, the literature on virtual patients and gamification \cite{ref5,ref6,ref12,ref13} suggests underexploited opportunities to integrate such technologies into massage therapy training. Furthermore, research on medical simulations that include advanced VR/AR and AI-driven systems demonstrates both the promise and the challenges of transferring these methods to fields where tactile interaction is essential \cite{ref11,ref20,ref21,ref22}.

\vskip 2mm
\noindent \textbf{Our Work vs. Others.} To these challenges and the gap in tactile-focused training identified above, we propose \textbf{TheraQuest}, a gamified web-based tool designed to immerse aspiring massage therapists in realistic clinical dialogues and decision-making scenarios. Using large language models to generate diverse patient profiles and employing a game engine for comprehensive assessment, TheraQuest enhances both the communicative and technical skills essential to successful massage therapy practice. Notably, it remains trainee-friendly and cost-effective, addressing the practical constraints often associated with more expensive AR/VR technologies.

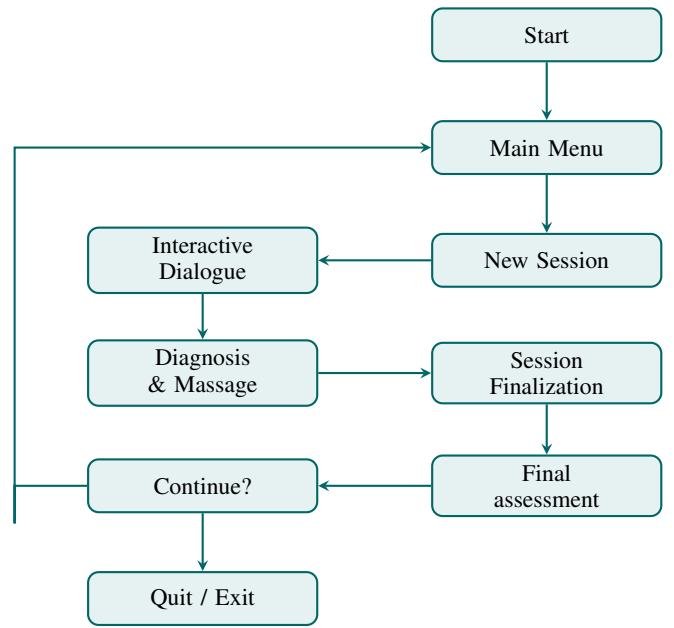
\begin{figure}[ht]
\centering
\begin{tikzpicture}[
    font=\small,
    node distance=1.5cm,
    every node/.style={align=center},
    block/.style={
        rectangle,
        thick,
        draw=teal!80!black,
        fill=teal!10,
        rounded corners,
        text width=8em,
        text centered,
        minimum height=2em
    },
    arrow/.style={
        ->,
        >=stealth,
        thick,
        draw=teal!80!black
    }
]

% Nodes
\node[block] (start) {Start};
\node[block, below of=start] (menu) {Main Menu};
\node[block, below of=menu] (newsession) {New Session};
\node[block, left=1.5cm of newsession] (dialogue) {Interactive \\ Dialogue};
\node[block, below of=dialogue] (diagnosis) {Diagnosis \\ \& Massage};
\node[block, right=1.5cm of diagnosis] (checkin) {Session \\ Finalization};
\node[block, below of=checkin] (assessment) {Final \\ assessment};
\node[block, left=1.5cm of assessment] (continue) {Continue?};
\node[block, below of=continue] (end) {Quit / Exit};

% Arrows
\draw[arrow] (start) -- (menu);
\draw[arrow] (menu) -- (newsession);
\draw[arrow] (newsession) -- (dialogue);
\draw[arrow] (dialogue) -- (diagnosis);
\draw[arrow] (diagnosis) -- (checkin);
\draw[arrow] (checkin) -- (assessment);
\draw[arrow] (assessment) -- (continue);
\draw[arrow] (continue) -- (end);
\draw[arrow] (continue) -| ++(-2.5, -0.5) |- (menu);

\end{tikzpicture}
\caption{Flowchart of the TheraQuest Prototype System Design (Version: 02/14/2025). 
trainees begin by registering or logging in [Start], then proceed to the [Main Menu] to either review their [User Profile], [Global Rankings] or [Therapy Room]. 
If the [New Session] is chosen, the trainee explores the patient's symptoms, leading to the [Diagnosis \& Massage] step. 
In [Session Finalization] step any additional follow-up communication or adjustments occur. The [Final Assessment] then shows the trainee's performance and potential improvement. Finally, the trainee decides whether to 
``Continue?'' (looping back to the main menu) or ``Quit / Exit.''}
\label{fig:TheraQuestFlowChart}
\end{figure}

\section{System Design}
In this section, we describe the core design and operation of the first prototype of TheraQuest. We start with trainee registration and entry into the main menu, followed by the creation of a virtual patient and an interactive dialogue where trainees diagnose problems and select appropriate massage techniques. We then outline the assessment loops, skill metrics, and ranking system that gamify the learning process, encouraging trainees to refine their therapeutic competencies and communication skills through repeated sessions. 

\subsection{Implementations}
The prototype version was developed using Microsoft C\# .NET Core and Razor 8.0 as the framework, while DALL-E, OpenAI, Unsplash \cite{ref32} and Behance \cite{ref31} was used to generate and search images.

\subsection{Main Menu}
TheraQuest first checks whether a trainee is already registered. If not, the system prompts them to create a new account by providing basic personal details and preferences. Returning trainees simply log in to retrieve their existing profiles. This step ensures that trainee data remains persistent across multiple sessions.

\begin{figure}
    \centering
    \includegraphics[width=0.95\linewidth]{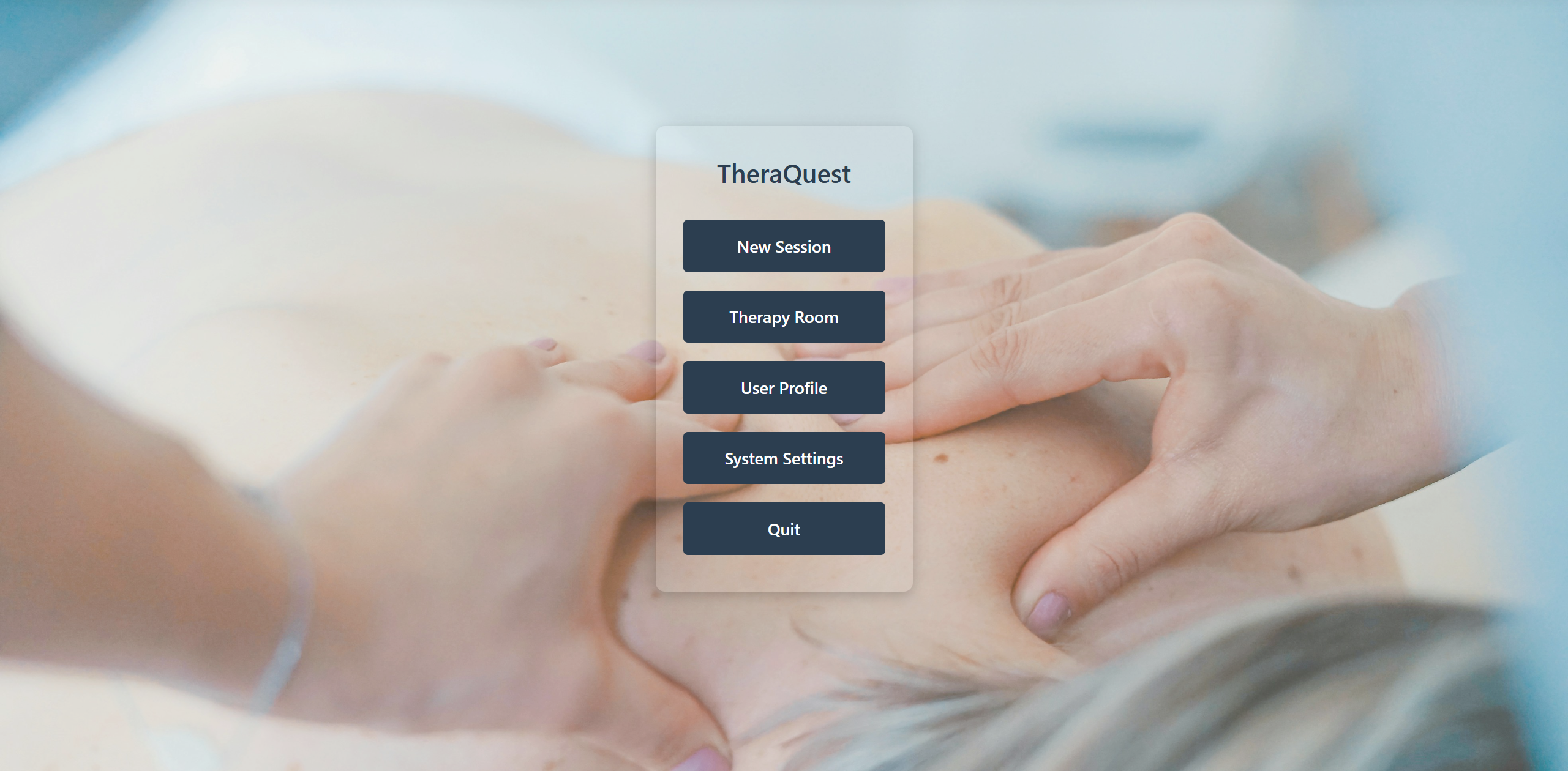}
    \caption{Main menu of the TheraQuest prototype. The background image shows the default massage room setting generated by DALL-E, OpenAI. Users can customize room backgrounds.}
    \label{img:mainMenu}
\end{figure}

Next, trainees see the main menu. From here, they can access \textbf{New Session} (to start a new patient session), \textbf{Therapy Room} (adjust virtual furniture, music, or ambiance), \textbf{Trainee Profile} (view skill progress, ranks, earned stars, and compare scores with other trainees), and \textbf{System Settings} (e.g., adjust audio volume or language preferences). A dedicated \textbf{Quit} button allows trainees to exit the session safely. The flow chart of the TheraQuest is shown in Figure \ref{fig:TheraQuestFlowChart}, and a sample main menu is shown in Figure~\ref{img:mainMenu}.

\subsection{New Session}

\begin{figure}
    \centering
    \includegraphics[width=0.95\linewidth]{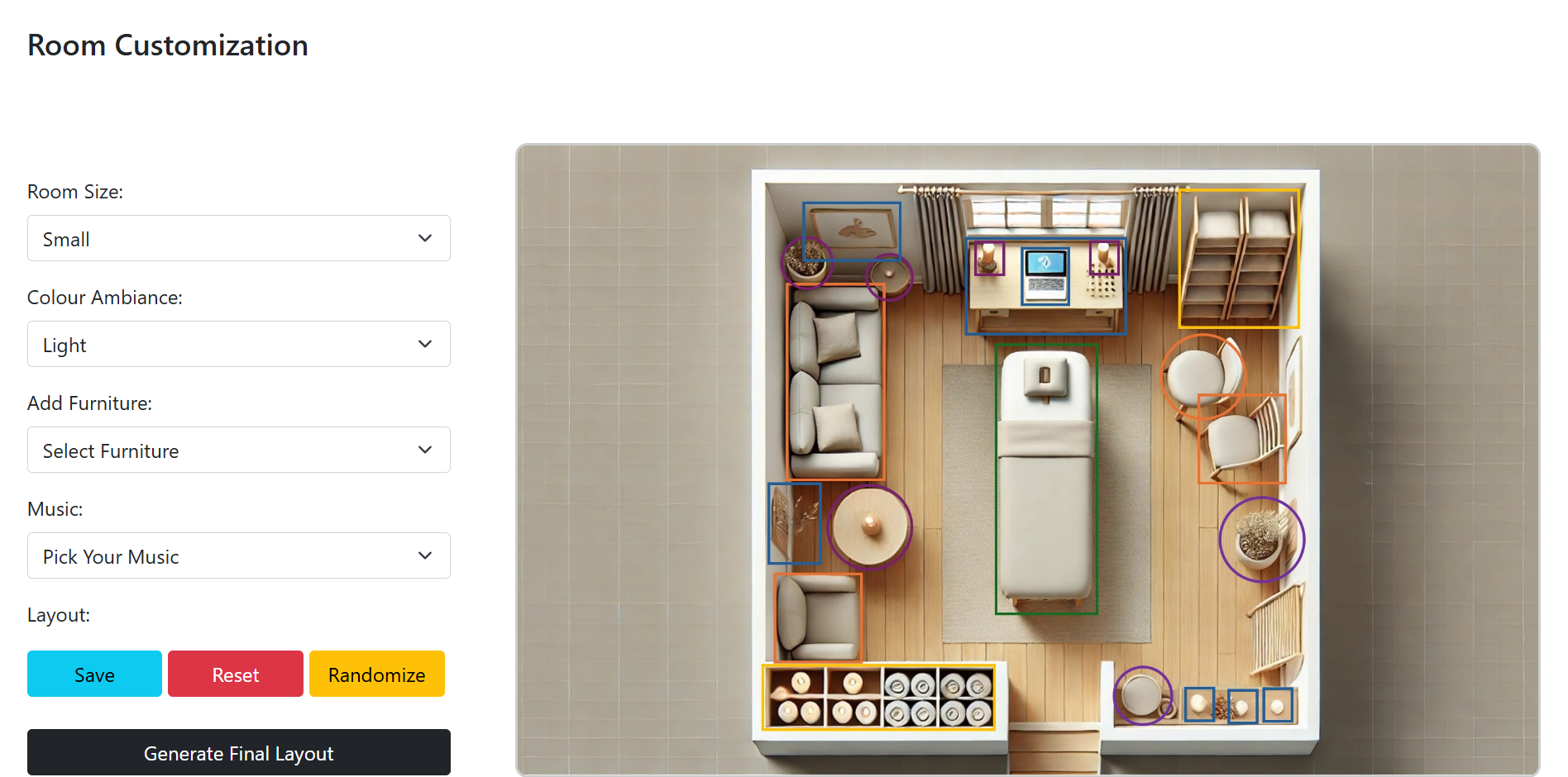}
    \caption{Trainees can adjust room size, colour ambiance, and furniture layout through an interactive floor plan, with options to save, reset, or randomize the setup. The final customized floor plan will be rendered into a 3D visualization for enhanced spatial representation.}
    \label{fig:MassageRoom}
\end{figure}

\vskip 1mm
\noindent \textbf{Therapy Room.} If the trainee has not customized their therapy room, a default floor plan is provided. The therapy room allows trainees change aspects such as room size, furniture placement, light ambience, and background music, helping to simulate a comfortable clinical environment and provide them with innovative ideas for their real massage room settings. Finally, the floor plan will be rendered in 3D and display as the background of the main menu. A screenshot for the therapy room is shown in Figure \ref{fig:MassageRoom}.

\vskip 1mm
\noindent \textbf{Interactive Dialogue.} When trainees begin a new therapy scenario, TheraQuest generates a virtual patient with a hidden symptom profile, ask the trainee to actively engage in diagnostic questioning. Unlike traditionally trained models that provide predefined case studies with fixed options and explicit symptoms, TheraQuest employs an AI-driven adaptive simulation that requires trainees to use a text box or microphone to communicate with patients, creating a dynamic and interactive learning experience.

\begin{figure}
    \centering
    \includegraphics[width=0.95\linewidth]{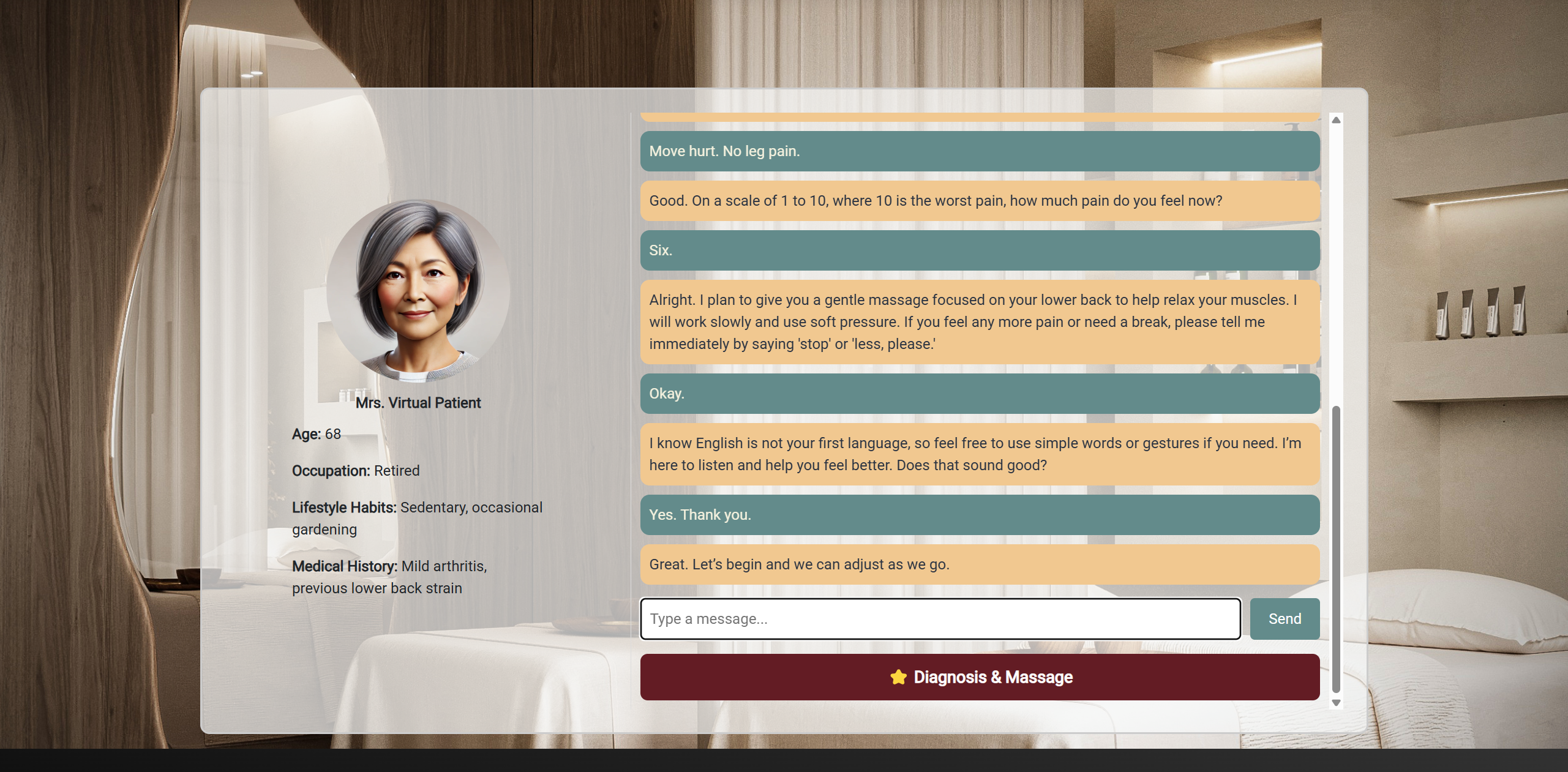}
    \caption{An AI-generated virtual patient, displaying the patient’s demographic details, lifestyle, and medical history is shown on the left side. Trainees interact via a live chat window on the right side to gather symptoms and provide instructions before proceeding to the ``Diagnosis \& Massage'' stage for hands-on decision-making.}
    \label{fig:conversation}
\end{figure}

The system first presents a brief on-screen introduction featuring the patient's avatar or profile photo, along with contextual background details such as age, occupation, lifestyle habits, and medical history information. These elements serve as important indicators that guide the trainee to formulating targeted questions, mimicking real-world clinical settings where patients may not always articulate their concerns clearly. However, the specific complaint remains undisclosed, requiring the trainees to engage in open-ended questioning to uncover relevant information. A screenshot of a conversation is shown in Figure \ref{fig:conversation}.
At the core of TheraQuest’s virtual patient system is a fine-tuned GPT model designed to generate a diverse range of patient types with varying communication styles, pain descriptions, and cultural backgrounds. This feature addresses a common limitation in healthcare training, which lack of exposure to diverse patient presentations. Assisted by natural language processing (NLP), TheraQuest enables virtual patients to respond dynamically to trainee input, ensuring that every interaction is unique. Some patients may be highly descriptive, clearly describing their pain points, while others may be vague, requiring the trainee to considerate deeper with follow-up questions. This settings trains trainees to navigate different patient personalities and communication barriers, preparing them for real-world complexities.

In addition, virtual patients may suffer from factors such as stress, cognitive bias, or prior misdiagnosis, so they sometimes may show misleading or conflicting symptoms 
in their descriptions during the chat; this feature simulates real-life cases where patients may unintentionally provide inaccurate descriptions. Since we cannot predict future patient cases, TheraQuest ensures that virtual patient interactions remain unpredictable and contextually variable, reflecting the inherent uncertainties of real-world clinical scenarios. 

\begin{figure}
    \centering
    \includegraphics[width=0.95\linewidth]{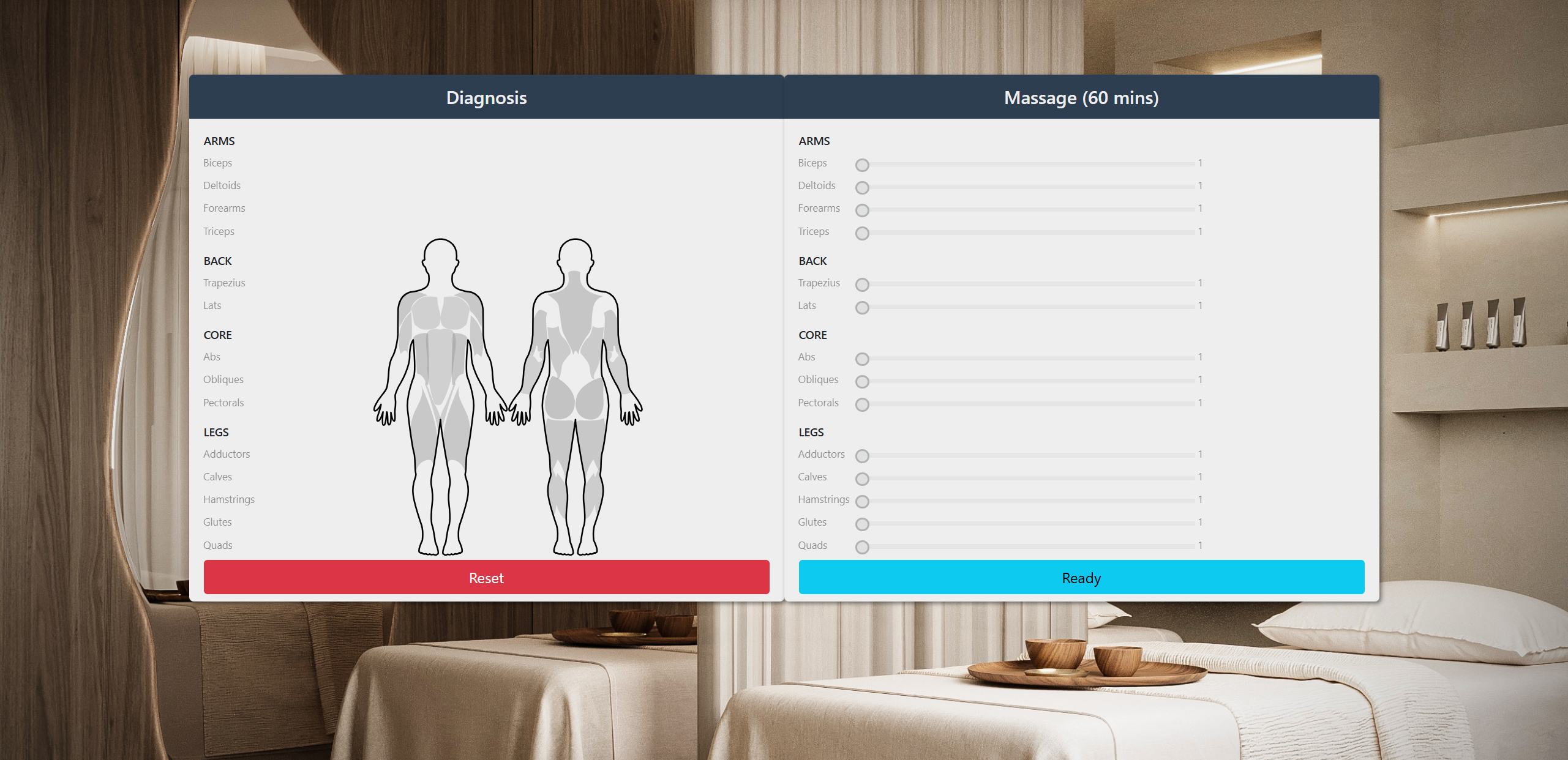}
    \caption{The left panel allows trainees to select muscle groups for diagnosis, and the right panel enables setting massage intensity through sliders. Disabled sliders appear in gray until their corresponding muscle groups are selected. The original muscle map comes from \cite{ref25}.}
    \label{img:bodyMap}
\end{figure}

\vskip 1mm
\noindent \textbf{Diagnosis and Massage.} After the conversation, the trainee attempts to diagnose the patient's problem area by selecting one or more regions on an interactive body map (shown in Figure~\ref{img:bodyMap}). Once the area(s) has been confirmed, the trainee chooses an appropriate massage technique and pressure level (drag the level bar up or down to choose a suitable pressure level(s)), then applies pressures virtually to the indicated areas. It is worth noting that the massage time is one hour (fixed value) in the prototype version.

\vskip 1mm
\noindent \textbf{Session Finalization.} Upon finishing the massage component, TheraQuest transitions to a post-massage check-in using the live chat window again, where the trainee converses again with the patient to gauge comfort and immediate relief. If the patient remains uncomfortable or reports unusual persistent pain, the trainee should adjust the technique or pressure level for future treatment. This iterative assessment loop emulates a real therapist's responsiveness, further enhancing the communication and dialogue metrics. 

\vskip 1mm
\noindent \textbf{Final assessment.} At the end of each session, a final assessment is used to evaluate the trainee's performance using different metrics. It shows achievements (e.g., effective questioning, correct pressure application) and suggests areas for improvement (e.g., missed follow-up questions, incorrect area(s) or pressure level(s)). By showing which specific actions influenced patient outcomes, TheraQuest promotes reflective learning and continuous growth.

\begin{figure}[h]
    \centering
    \includegraphics[width=0.95\linewidth]{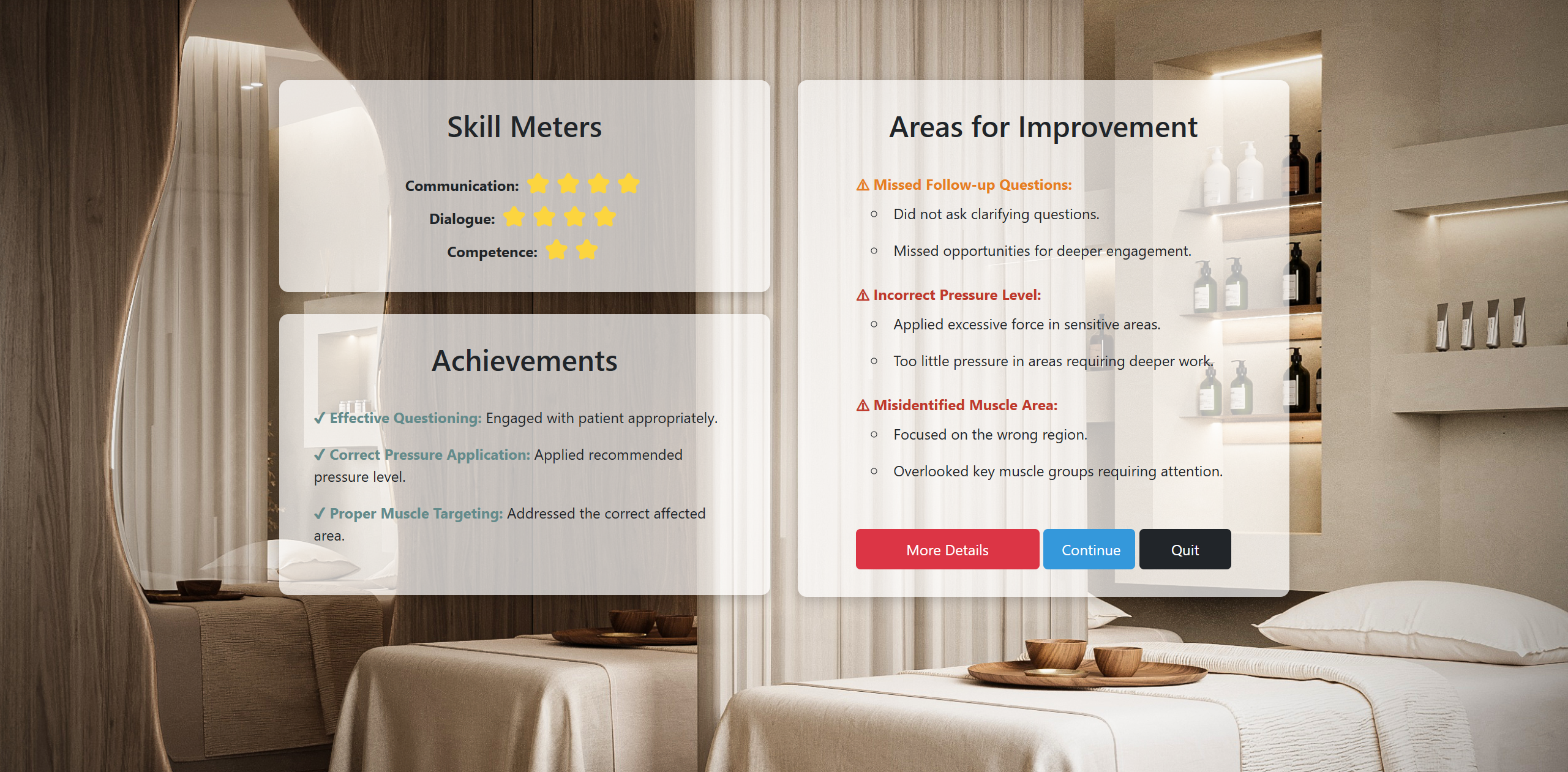}
    \caption{The final assessment page provides a structured evaluation of a trainee's performance after each session. The left panel displays skill meters for \textbf{Communication, Dialogue, and Competence}, where users earn \textbf{1-5 stars} based on their performance. Below the skill meters, an \textbf{Achievements} section lists successfully performed actions, such as effective questioning and correct pressure application. The right panel, \textbf{Areas for Improvement}, uses a color-coded system where \textcolor{orange}{orange} iullstrates moderate mistakes (e.g., missed follow-up questions) and \textcolor{red}{red} shows severe mistakes (e.g., incorrect pressure levels or misidentified muscle areas).}
    \label{fig:final_assessment}
\end{figure}

Finally, trainees can go back to the main menu or quit the application. Any progress made, including skill improvements, earned ``stars'', and session metrics (e.g., number of patients served, average satisfaction rating), are saved to the trainee's profile. This persistent record improves a sense of continuity and growth, encouraging trainees to return, refine their approach, and master increasingly complex scenarios. A screenshot of final assessment is shown in Figure \ref{fig:final_assessment}. 

\begin{figure}[h]
    \centering
    \includegraphics[width=0.95\linewidth]{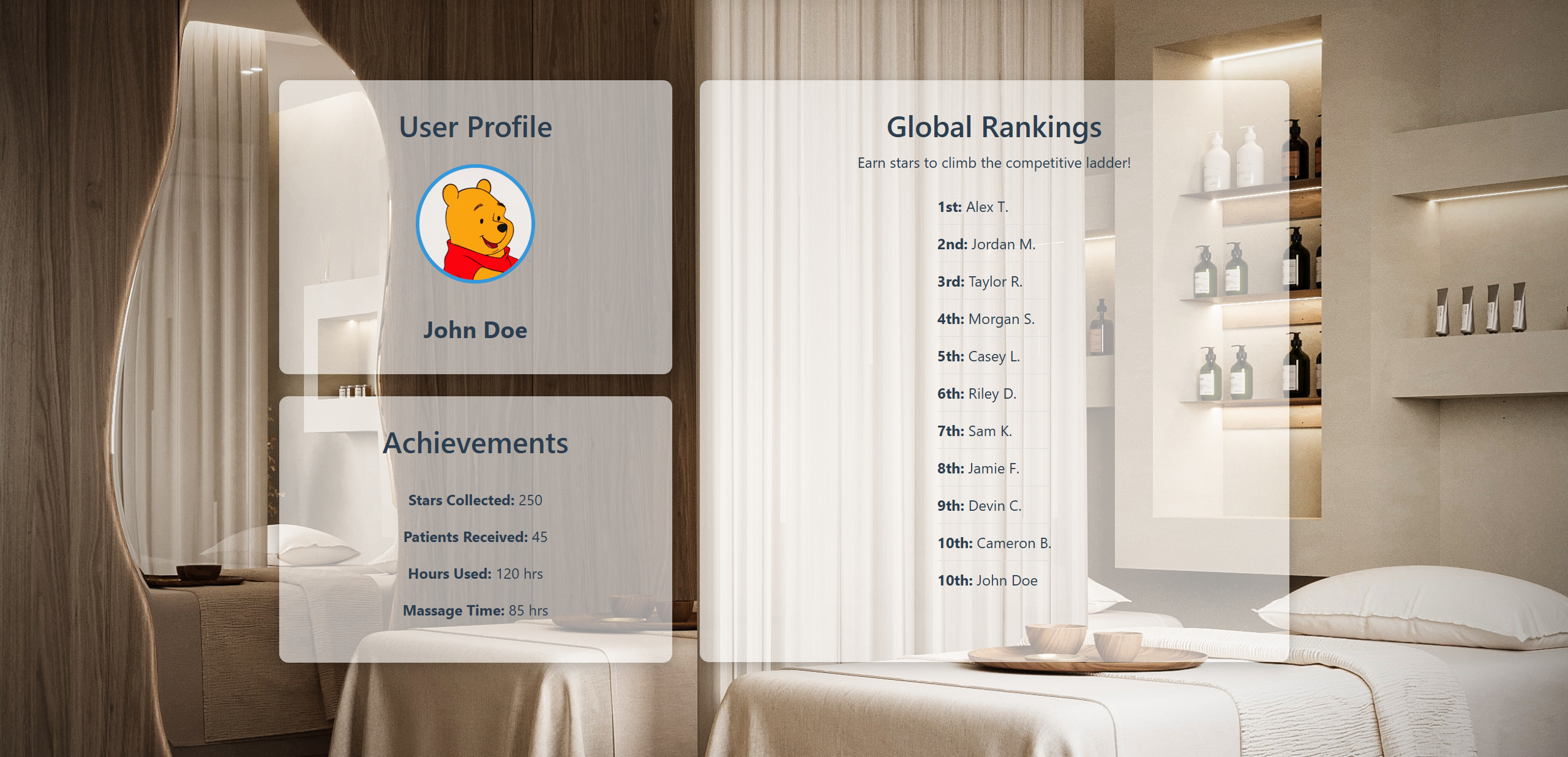}
    \caption{The user profile page displays the trainee's overall progress, skill meters, and ranking in different categories. On the left side, the trainee's avatar, name, collected stars, received patients, total usage hours, and massage time are shown. The right side presents the global ranking for Communication, Dialogue, Competence, and Overall Performance. The ranking system encourages trainees to improve their skills through a competitive leaderboard.}
    \label{fig:Profile}
\end{figure}

\vskip 1mm
\noindent \textbf{Trainee Profile} Using the data from the final summary, the trainee's skill meters (Communication, Dialogue, Competence) are updated to reflect overall progress. The system awards ``stars'' based on performance, which contribute to a global ranking leaderboard. This feature motivates trainees to refine their skills, accumulate stars, and climb the competitive ladder, making TheraQuest not only an educational platform but also an engaging and gamified experience.  A screenshot of the trainee profile is shown in Figure~\ref{fig:Profile}.

\section{Evaluations}
\label{sec:evaluations}

In this section, we propose a two-stage evaluation strategy for our TheraQuest prototype. Prior to the formal study, we will run multiple rounds of pilot tests with lab members to ensure that no serious bugs remain and that the study procedure is clear. First, we will collaborate with local professional massage therapists  to validate the realism and accuracy of the system. Then we will conduct a user study to examine TheraQuest’s effectiveness compared to traditional teaching methods (and other relevant approaches, if applicable).

\subsection{Expert Review}
Before testing with trainees, we will invite a group of (est. 5--10) local professional massage therapists to examine the prototype (in person) when there are many massage therapists working near our university. Their evaluation will focus on:
\begin{itemize}
    \item \textbf{AI-Generated Virtual Patients:} Assess the realism and accuracy of the symptoms and demographic profiles created by the LLM-driven module.
    \item \textbf{Final Assessment Completeness:} Confirm whether the metrics and feedback are clinically valid and cover the core competencies expected in massage therapy practice.
    \item \textbf{System Functionality and Workflow:} Identify any missing features, technical issues, or usability concerns from an expert perspective.
\end{itemize}
Feedback and interactive data from these expert reviews will be used to refine the prototype and address any gaps before the user study.

\subsection{User Study Design}
\noindent \textbf{Recruitment and Pre-study Questionnaire.}
Since our target participants are \emph{local massage therapy trainees}, online recruitment via common crowdsourcing platforms (e.g., Prolific or Amazon MTurk) would not be suitable any may have potential bots or puppeteers issues \cite{ref33}. Instead, we plan to conduct in-person user studies by recruiting students in local colleges. Participants will complete a brief survey on their baseline knowledge of massage therapy techniques and their opinions of massage therapy education. The actual number of invited participants will be calculated by G power, we estimate the minimum number of participants will be around 50 -- 80.

\vspace{1mm}
\noindent \textbf{Group Assignment.}
To test the effectiveness of our serious game, participants will be randomly divided into two groups:
\begin{enumerate}
    \item \textbf{Control Group:} Learns through traditional methods (e.g., lectures, standard readings, or supervised practice).
    \item \textbf{Simulation Group:} Learns using our TheraQuest prototype, which simulates interactive massage scenarios.
\end{enumerate}

\noindent We will employ \textbf{two different conditions} for comparison:

\begin{itemize}
    \item \textbf{Condition A (Fixed Time):} Both groups receive the same, fixed amount of training time, after which participants take a knowledge quiz. This allows us to evaluate whether the Simulation Group achieves higher knowledge gains \emph{within a given time frame}.
    \item \textbf{Condition B (Target Knowledge):} Both groups continue training until they achieve a predetermined proficiency score on practice quizzes. We then compare the total time each participant needed to reach this target. This measures whether the Simulation Group can \emph{reach the same competency level in less time}.
\end{itemize}

\subsection{Post-study Assessment}
After the learning sessions, all participants will complete a second questionnaire designed to:
\begin{enumerate}
    \item \textbf{Measure Knowledge Gain:} A quiz on fundamental massage therapy concepts to compare learning outcomes between the Control and Simulation groups under both Condition A and Condition B.
    \item \textbf{Attitude Change:} Participants’ confidence and perceptions about massage therapy training are measured before and after the intervention, allowing us to see whether the serious game yields more positive changes.
    \item \textbf{Usability \& Engagement:} Since our target audience is \emph{massage therapy trainees}, we will use a short questionnaire (rather than a formal SUS scale) to measure enjoyment of the game, ease of use, perceived relevance, and collect any open-ended comments.
    \item \textbf{Time Efficiency:} In Condition B, we will record the time it takes for each trainee to reach a defined competency level and compare the results between the two groups.
\end{enumerate}

\section{Results} 
We will use independent t test to investigate the significance. We hypothesize that the \emph{Simulation Group} will show greater improvements in communication skills, theoretical and pratical knowledge (Condition A) and require less time to reach a given competency level (Condition B), while also reporting more favourable attitudes toward massage therapy training. 

\section{Discussions}

In this section, we evaluate the current TheraQuest prototype, address its limitations, and discuss major challenges and future development directions.

\subsection{Limitations}
TheraQuest is still in the early prototype stage and currently focuses only on massage therapy training, limiting its applicability to other healthcare disciplines such as physiotherapy or chiropractic medicine. In addition, some advanced massage techniques and complex patient conditions are not fully integrated, reducing the depth of simulated scenarios. The web-based nature of the system, while convenient for broad accessibility, lacks the immersive potential of virtual or augmented reality. While VR/AR could enhance realism, it also introduces constraints such as hardware costs and trainee comfort issues. Since TheraQuest has not yet undergone extensive empirical evaluations or randomized controlled trials, its effectiveness currently relies on preliminary observations, which may introduce bias.

\subsection{Major Challenges}
There are several key challenges that must be addressed. The first challenge is properly training LLMs with domain-specific data to generate realistic virtual patients when very limited datasets can be used. Another related challenge is embedding high-quality domain knowledge into the system. Massage therapy is a hands-on tactile practice that goes far beyond text descriptions. Designing rich clinical logic, such as recommended pressure levels, and generating professional assessment requires close collaboration with domain experts to ensure the system presents accurate, evidence-based results.  

Another important challenge concerns the experience of the trainees and sustained participation. Although the gamified elements of TheraQuest, such as skill metrics and star-based achievements, can motivate learners, too many game mechanics or a poorly structured interface risk confusing or overwhelming trainees. In contrast, overly simplistic interactions may not challenge trainees who already have advanced clinical skills. Achieving a balance that accommodates diverse learner profiles, from novices to seasoned practitioners, is an ongoing goal that will require continuous training and assessment loops of trainees.

\subsection{Future Work}
Moving forward, TheraQuest can be strengthened through multiple approaches. One of the most pressing needs is a complete set of user studies that involve both students and professional massage therapists, providing empirical data and feedback on how the system affects clinical reasoning, interpersonal communication, and practical competencies. Collaboration with certified practitioners and massage therapy educators will enable more vital virtual patient scenarios, as well as more precise tuning of recommended techniques and pressure levels.  

Another direction is to expand TheraQuest's functionality and scope. Adaptations could include specialized modules for physical therapy and rehabilitation, each reflecting the unique demands of different patient populations. Although the system currently using a web-based approach, future prototypes might introduce optional VR or AR components for those seeking enhanced immersion. Incorporating features like multi-session patient tracking would allow trainees to observe progress over multiple visits, mirroring real-world practice where conditions evolve over time. By following these paths, TheraQuest has the potential to become a robust and widely adopted platform that effectively leverages LLM technology for interactive, skill-based healthcare training.

\section{Conclusion}
We designed TheraQuest, a novel gamified, LLM-driven platform to bridge the communication gap in massage therapy training by simulating realistic patient-therapist interactions through dialogue, diagnostic processes, and immediate assessment. Unlike many existing tools that focus on other healthcare fields or static tactile-based scenarios, TheraQuest targets both technical and interpersonal skills vital to massage therapy, allowing trainees to refine diagnostic reasoning and empathetic engagement in a user-friendly, web-based environment. Although TheraQuest still in the prototype stage, the customized therapy room, virtual patients generated from diverse backgrounds, and skill metrics show potential to improve learning outcomes, especially when further validated through collaboration with professional massage therapists and larger-scale user studies.

\end{document}